# The resting microstate networks (RMN): cortical distributions, dynamics, and frequency specific information flow

Roberto D Pascual-Marqui[1,2], Dietrich Lehmann[1], Pascal Faber[1], Patricia Milz[1], Kieko Kochi[1], Masafumi Yoshimura[2], Keiichiro Nishida[2], Toshiaki Isotani[3], Toshihiko Kinoshita[2]

[1]The KEY Institute for Brain-Mind Research, University of Zurich, Switzerland
[2]Department of Neuropsychiatry, Kansai Medical University, Osaka, Japan
[3]Laboratory for Brain-Mind Research, Shikoku University, Tokushima, Japan

## 1. Abstract

A brain microstate is characterized by a unique, fixed spatial distribution of electrically active neurons with time varying amplitude. It is hypothesized that a microstate implements a functional/physiological state of the brain during which specific neural computations are performed. Based on this hypothesis, brain electrical activity is modeled as a time sequence of non-overlapping microstates with variable, finite durations (Lehmann and Skrandies 1980, 1984; Lehmann et al 1987). In this study, EEG recordings from 109 participants during eyes closed resting condition are modeled with four microstates. In a first part, a new confirmatory statistics method is introduced for the determination of the cortical distributions of electric neuronal activity that generate each microstate. All microstates have common posterior cingulate generators, while three microstates additionally include activity in the left occipital/parietal, right occipital/parietal, and anterior cingulate cortices. This appears to be a fragmented version of the metabolically (PET/fMRI) computed default mode network (DMN), supporting the notion that these four regions activate sequentially at high time resolution, and that slow metabolic imaging corresponds to a low-pass filtered version. In the second part of this study, the microstate amplitude time series are used as the basis for estimating the strength, directionality, and spectral characteristics (i.e., which oscillations are preferentially transmitted) of the connections that are mediated by the microstate transitions. The results show that the posterior cingulate is an important hub, sending alpha and beta oscillatory information to all other microstate generator regions. Interestingly, beyond alpha, beta oscillations are essential in the maintenance of the brain during resting state.

## 2. Introduction

A brain microstate is characterized by a unique, fixed spatial distribution of electrically active neurons with time varying amplitude. It is hypothesized that a microstate implements a functional/physiological state of the brain during which specific neural computations are performed. Based on this hypothesis, brain electrical activity is modeled as a time sequence of non-overlapping microstates with variable finite duration (Lehmann and Skrandies 1980, 1984; Lehmann et al 1987).

In this study, the human brain in the awake, eyes closed, resting state (abbreviated as EC REST) will be studied, using recorded EEGs from a large number of healthy participants. Using the microstate model as the basis of the study, the aims are:
1. To determine the cortical distribution of electrically active neurons that generates each microstate.
2. To model the time varying properties of the microstates and on this basis to determine the cortical functional connectivity that ensues.

## 3. Material

Real human EEG recordings in the awake, eyes closed, resting state (EC REST), using 64-channel EEG recordings from 109 participants, are publicly available from Goldberger et al. (2000), Schalk et al. (2004). Each recording consists of 1 min. EEG, sampled at 160 Hz. Three electrodes (T9, T10, and Iz) were discarded for analysis, because they were spatial outliers relative to the other 61 electrodes that cover the scalp in an approximate uniformly distributed manner.

The LORETA-KEY software was used for all analyses. This is free academic software, available online at:
www.uzh.ch/keyinst/loreta

## 4. Methods: General

A unique mathematical expression for the microstate model for EEG measurements can be written as:

**Eq. 1** $\quad \mathbf{\Phi}_t = \mathbf{\Gamma} \mathbf{a}_t$

where $\mathbf{\Phi}_t \in \mathbb{R}^{N_E \times 1}$ denotes an instantaneous (at discrete time sample "$t$") EEG average reference scalp map (i.e. scalp electric potential differences null sum over all electrodes) measured at $N_E$ electrodes; $\mathbf{\Gamma} \in \mathbb{R}^{N_E \times N_\mu}$ denotes $N_\mu$ microstate scalp maps, contained in the columns of $\mathbf{\Gamma}$, which are normalized (e.g. they have unit global field power [GFP=1]); and $\mathbf{a}_t \in \mathbb{R}^{N_\mu \times 1}$ contains the microstate amplitudes at time "$t$".

The microstate model is defined under the constraint that there is only one non-null element in $\mathbf{a}_t$ at any given moment in time. Under these conditions, and if the number of microstates $N_\mu$ is given, all parameters ($\mathbf{\Gamma}$, and $\mathbf{a}_t \ \forall t$) can be estimated by the method and algorithm described in detail in (Pascual-Marqui et al 1995). For convenience, a "*label time series*" is introduced, denoted as $L_t$, which takes integer values in the range $1...N_\mu$, indicating which microstate is active, i.e. which element of $\mathbf{a}_t$ is non-zero.

Essentially, the algorithm iterates 2 main steps:
Step#1. Given the normalized microstates scalp maps $\left( \mathbf{\Gamma}_1, \mathbf{\Gamma}_2, ..., \mathbf{\Gamma}_{N_\mu} \right)$. Then for each measured EEG scalp map $\mathbf{\Phi}_t$, compute:

**Eq. 2** $\quad b_{it} = \mathbf{\Phi}_t^T \mathbf{\Gamma}_i$

where the superscript "$T$" denotes vector-matrix transpose. Then find which microstate it belongs to:

**Eq. 3** $\quad L_t = \arg \max_i b_{it}^2$

Then set all elements of $\mathbf{a}_t$ to zero, except for the κ-th element, with:

**Eq. 4** $\quad \kappa = L_t$

which corresponds to its microstate, i.e.:

**Eq. 5** $\quad \left[ \mathbf{a}_t \right]_\kappa = b_{\kappa t}$

Step#2. Given the labels $L_t$ for each measured EEG scalp map $\mathbf{\Phi}_t$, compute the new normalized microstates scalp maps as the first eigenvector of the covariance matrix of the set of measured EEG scalp maps for each label value separately. For instance:

Eq. 6
$$\begin{cases} \Gamma_i = \arg\max_{\mathbf{V}} \left[ \mathbf{V}^T \left( \sum_{\forall t, s.t.: L_t = i} \Phi_t \Phi_t^T \right) \mathbf{V} \right] \\ s.t.: \mathbf{V}^T \mathbf{V} = 1 \end{cases}$$

The detailed derivation of the algorithm can be found in (Pascual-Marqui et al 1995).

Now consider the following problem:
Given the microstate scalp maps $\Gamma \in \mathbb{R}^{N_E \times N_\mu}$ and a measured EEG scalp map $\Phi_t \in \mathbb{R}^{N_E \times 1}$, find the least squares estimator for the model:

Eq. 7    $\Phi_t = \Gamma \mathbf{c}_t$

without constraints on the unknown coefficients $\mathbf{c}_t \in \mathbb{R}^{N_\mu \times 1}$.

Note that this problem deviates from the strict definition of the microstate model, where only one microstate scalp map is active at any time. In the unrestricted model of Eq. 7, the microstate scalp maps will combine linearly in order to explain (i.e. to fit) the measured data $\Phi_t$.

The solution to the least squares problem in Eq. 7 is:

Eq. 8    $\mathbf{c}_t = \left( \Gamma^T \Gamma \right)^{-1} \Gamma^T \Phi_t$

This solution has the following trivial property:
If a measured EEG scalp map $\Phi_t$ exactly fits the strict microstate model, i.e. one and only one the microstate scalp maps in the columns of $\Gamma$ is exactly collinear with $\Phi_t$, then all coefficients in $\mathbf{c}_t$ are zero, except for the one corresponding to its generative microstate scalp map.

What this means is that if the measured EEG can be represented to a high degree of accuracy with a microstate model, then the coefficients $\mathbf{c}_t$ obtained from Eq. 8 are almost identical to the strict microstate coefficients $\mathbf{a}_t$ obtained from Step#1 of the algorithm described above by Eq. 2, Eq. 3, Eq. 4, and Eq. 5.

Within the time intervals in which a microstate is stable, all measured EEG scalp maps are approximately collinear to the corresponding microstate scalp map. In the time period around a microstate transition, the cortical generators are changing their spatial distribution, and the measured EEG scalp maps are not collinear. These transitions are well captured by the relaxed (non-constrained) estimation in Eq. 8.

We will therefore denote $\mathbf{c}_t$ (Eq. 8) as the "*microstate time series*", in distinction to the "*label time series*" $L_t$ (Eq. 3).

Traditionally, most previous microstate publications characterize the dynamics of the model with features extracted from the *label time series* $L_t$. Examples of such features consist of the average time spent in each microstate, the mean duration of the microstate intervals, the transitions using Markov theory, etc (see, for example, Lehmann et al 2005).

As an additional, different, new, and complimentary method, the *microstate time series* $\mathbf{c}_t$ (Eq. 8) will be used in this study, in order to obtain a characterization of the dynamics that will allow for determining the frequency specific cortical information flow.

In particular, the *microstate time series* $\mathbf{c}_t$ will be modeled as a multivariate autoregression, from which the isolated effective coherences (iCoh) can be estimated (Pascual-Marqui et al 2014). The isolated effective coherence can be used to assess the properties of the network formed by the set of different microstates, i.e., the strength, directionality, and spectral characteristics (i.e., which oscillations are preferentially transmitted) of the connections between the different microstate generators.

The detailed derivation of the isolated effective coherence (iCoh) can be found in (Pascual-Marqui et al 2014), together with links to software code and data that validate the method.

The other question of interest in this study is to determine the cortical distributions of electric neuronal activity that generate each microstate. For this purpose, "exact low resolution electromagnetic tomography" (eLORETA) (Pascual-Marqui 2007, Pascual-Marqui et al 2011) will be applied to microstate scalp maps. The literature offers a wide range of choice of inverse solutions to the EEG problem, some of which are reviewed in Pascual-Marqui (2009). The eLORETA method is used in this due to its unique property, not shared by any other linear inverse solution: exact localization to arbitrarily placed point sources. Based on the principles of linearity and superposition, this guarantees that eLORETA localizes correctly an arbitrary distribution, albeit with low spatial resolution. Furthermore, eLORETA is an improvement over the previous versions of the method, namely LORETA and standardized LORETA (sLORETA), both of which have been abundantly validated, as reviewed in (Pascual-Marqui et al 2011).

## 5. Methods: Applied to the 109-EEG data set

The EEGs from 109 participants, as described above in the "Material" section, were analyzed in the following way.

Firstly, the EEGs were digitally band-pass filtered using the discrete Fourier transform to the interval 2 to 20 Hz. This is the most common frequency range used in previous microstate analyses (see e.g. Lehmann et al 2005).

Next, the total average global field for each participant's 1-minute recording was calculated, and its inverse used as a single scalar factor applied to the participant's 1-minute recording, in order to achieve that all participants were comparable in terms of the average EEG amplitude. This scaling (i.e. this normalization) eliminates a source of variance which is not relevant to the microstate analysis. In this way, all the EEG data can be used together, to find the microstates common to all participants, while at the same time it is guaranteed that all participants contribute equally to the estimators.

This means that the total number of measured EEG scalp maps $\Phi_{jt}$ used in estimating four microstates, was 109 X 60 X 160, corresponding to "number of participants" X "number of seconds in one minute" X "the number of samples per minute", in other words, the subscripts in $\Phi_{jt}$ are in the ranges $j=1...N_S$, with $N_S=109$ (number of participants); and $t=1...N_T$, with $N_T=60\times 160=9600$.

All these maps were given to the algorithm outlined in Eq. 2 through Eq. 6, in order to estimate four common microstate scalp maps, denoted as $\Gamma_1, \Gamma_2, \Gamma_3, \Gamma_4 \in \mathbb{R}^{N_E \times 1}$. The choice of four microstates ($N_\mu = 4$) corresponds to the minimum cross-validation error in many EEG samples (see e.g. Koenig et al 2002). Note that this process also produces estimators for the *label time series* $L_{jt}$ for each individual participant. Given the *label time series* $L_{jt}$ for each individual participant, the individual microstate scalp maps $\Gamma_{j1}, \Gamma_{j2}, \Gamma_{j3}, \Gamma_{j4} \in \mathbb{R}^{N_E \times 1}$ can be computed using Eq. 6 in the following form:

Eq. 9
$$\begin{cases} \Gamma_{ji} = \arg\max_{\mathbf{V}} \left[ \mathbf{V}^T \left( \sum_{\forall t, s.t.: L_{jt}=i} \mathbf{\Phi}_{jt} \mathbf{\Phi}_{jt}^T \right) \mathbf{V} \right] \\ s.t.: \mathbf{V}^T \mathbf{V} = 1, \text{ for } j = 1...N_s, i = 1...N_\mu, \text{ with } N_\mu = 4 \end{cases}$$

It is worth emphasizing that the method presented here for estimating each participant's microstate scalp maps is distinctly different from the methodology commonly used in other publications, such as in (Koenig et al 2002), in which:
"Microstate class topographies were computed individually and averaged across participants, using a permutation algorithm that maximized the common variance over participants".

In our case:
"Microstate class topographies were computed on the full sample (all participants), and the individual labels were used to compute the individual topographies"

Finally, the *microstate time series* $\mathbf{c}_{jt} \in \mathbb{R}^{N_\mu \times 1}$ for each participant is computed from Eq. 8 as:

Eq. 10  $\mathbf{c}_{jt} = \left( \mathbf{\Gamma}^T \mathbf{\Gamma} \right)^{-1} \mathbf{\Gamma}^T \mathbf{\Phi}_{jt}$

## 6. Results and discussion

### 6a: The four common microstates

The four common microstate scalp maps $\Gamma_1, \Gamma_2, \Gamma_3, \Gamma_4 \in \mathbb{R}^{N_E \times 1}$ are shown in Figure 1a. These results are in fairly good qualitative agreement with four common microstate scalp maps obtained by Koenig et al 2002, using 19-channel EEG data from 496 participants. It should be noted that the correspondence observed here is based on the fact that EEG microstates are polarity invariant, because the neuronal generator distribution is the same regardless of the polarity of the map.

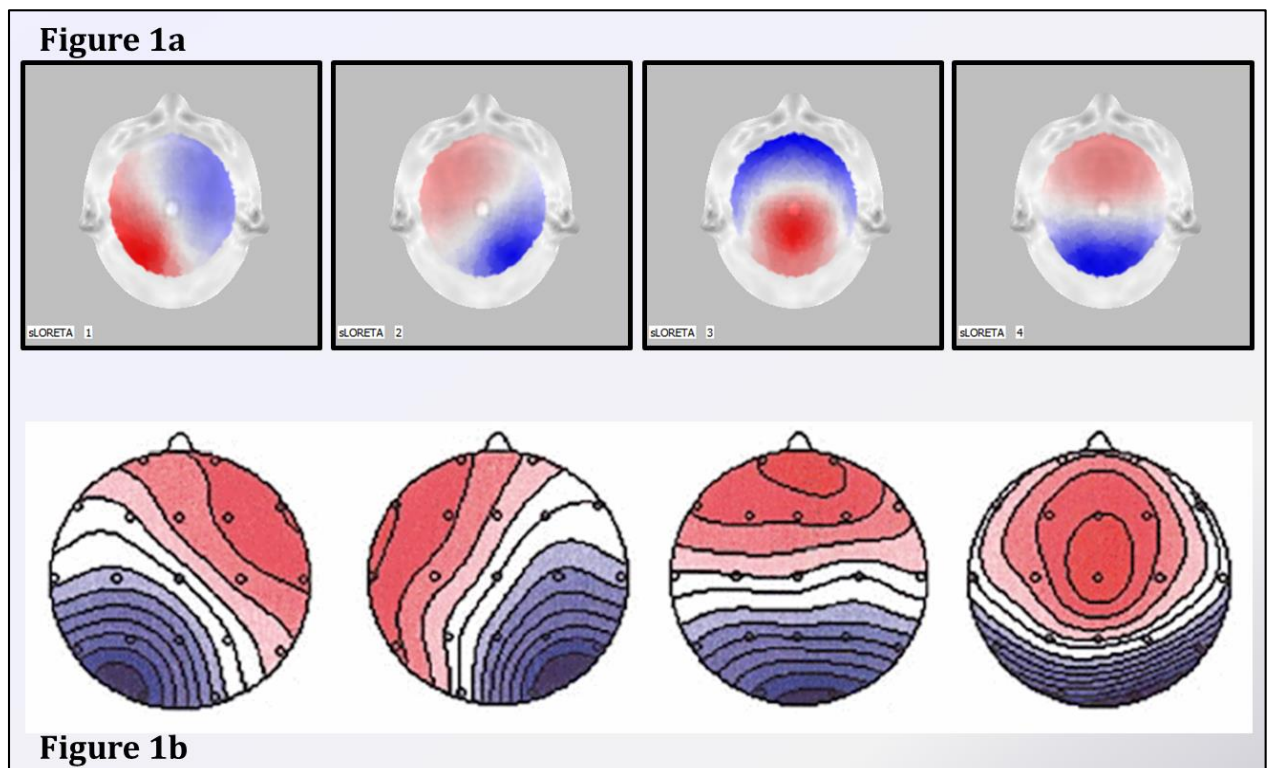

Figure 1: Four microstate scalp maps. 1a: The four common microstate scalp maps for 109 participants, this study. 1b: Four common microstate scalp maps for 496 participants, from (Koenig et al 2002). Color code: red for positive electric potential values, blue for negative. It should be noted that EEG microstates are polarity invariant, because the neuronal generator distribution is the same regardless of the polarity of the map.

## 6b: The cortical distribution of electrically active neurons underlying the four microstates: exploratory analysis

eLORETA computations can be directly applied to the four common microstate scalp maps. This will produce a functional mapping of the generators corresponding to eigenvector scalp maps. These 3D cortical images do not take into account the variance. They are shown in Figure 2.

Figure 2 shows that all microstates have a common strong generator in the posterior cingulate cortex. In addition, microstate 1 has maximum activity in left occipital, microstate 2 has maximum in right occipital, and microstate 3 in anterior cingulate areas.

These results are compatible with the notion that the cortical generators of the microstates correspond to a fragmented version of the metabolically determined default mode network (DMN) (see e.g. Raichle et al 2001, Greicius et al 2003). Thus, these results seem to indicate that metabolically determined DMN actually corresponds to a very low-pass time filtered version of the faster microstate dynamics.

In other words: Slow metabolic imaging blurs over time the actual elementary processes (metaphorically speaking: "atoms") that have much faster dynamics.

This is illustrated in Figure 3.

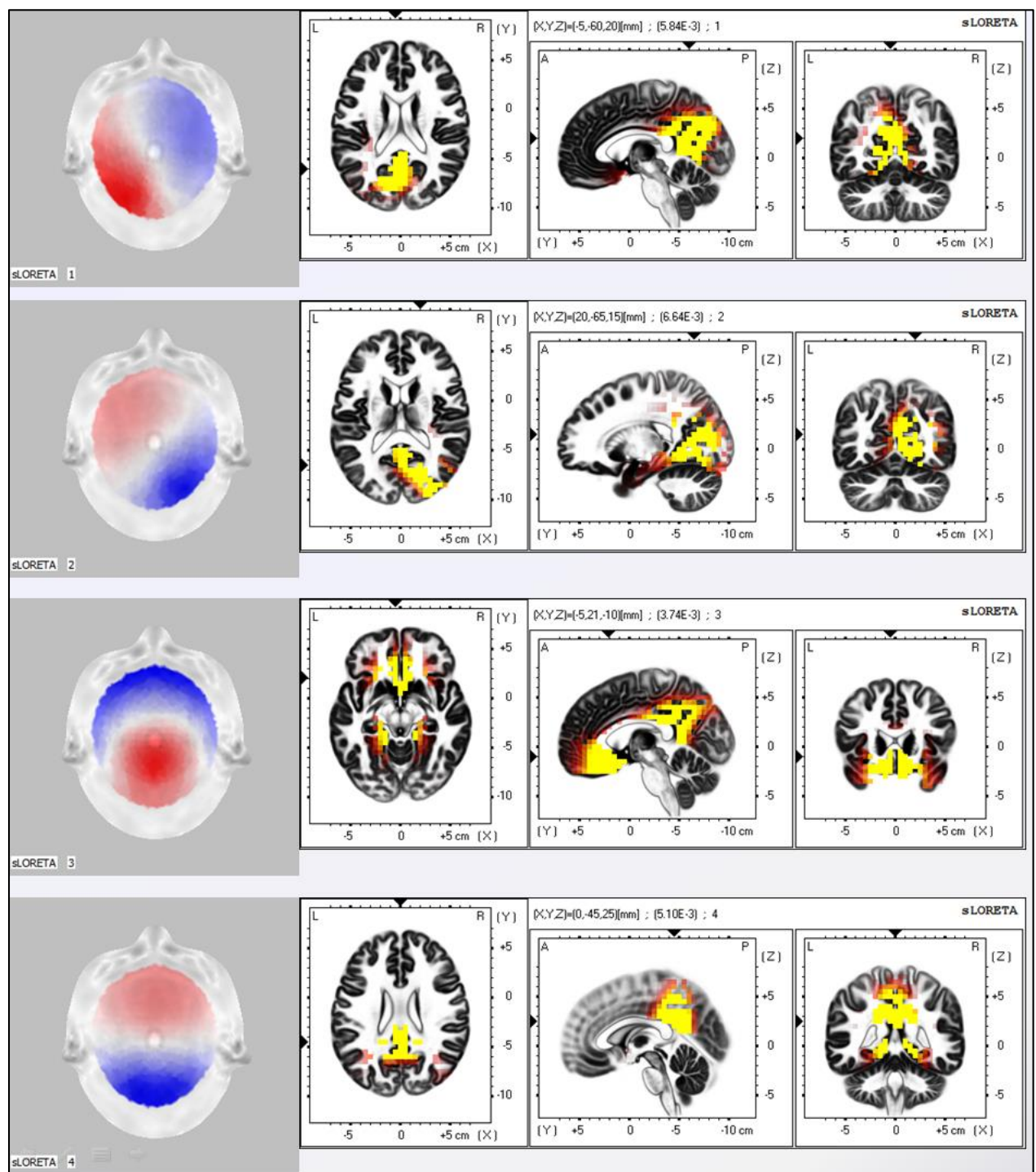

Figure 2: The four microstate scalp maps and their corresponding eLORETA images of electric neuronal activity, with bright yellow color encoding the maximum current density. In the eLORETA images, the two small triangles along the coordinate axes point to the maximum activity. Note that all microstates have a common strong generator in the posterior cingulate cortex. In addition, microstate 1 has maximum activity in left occipital, microstate 2 has maximum in right occipital, and microstate 3 in anterior cingulate areas.

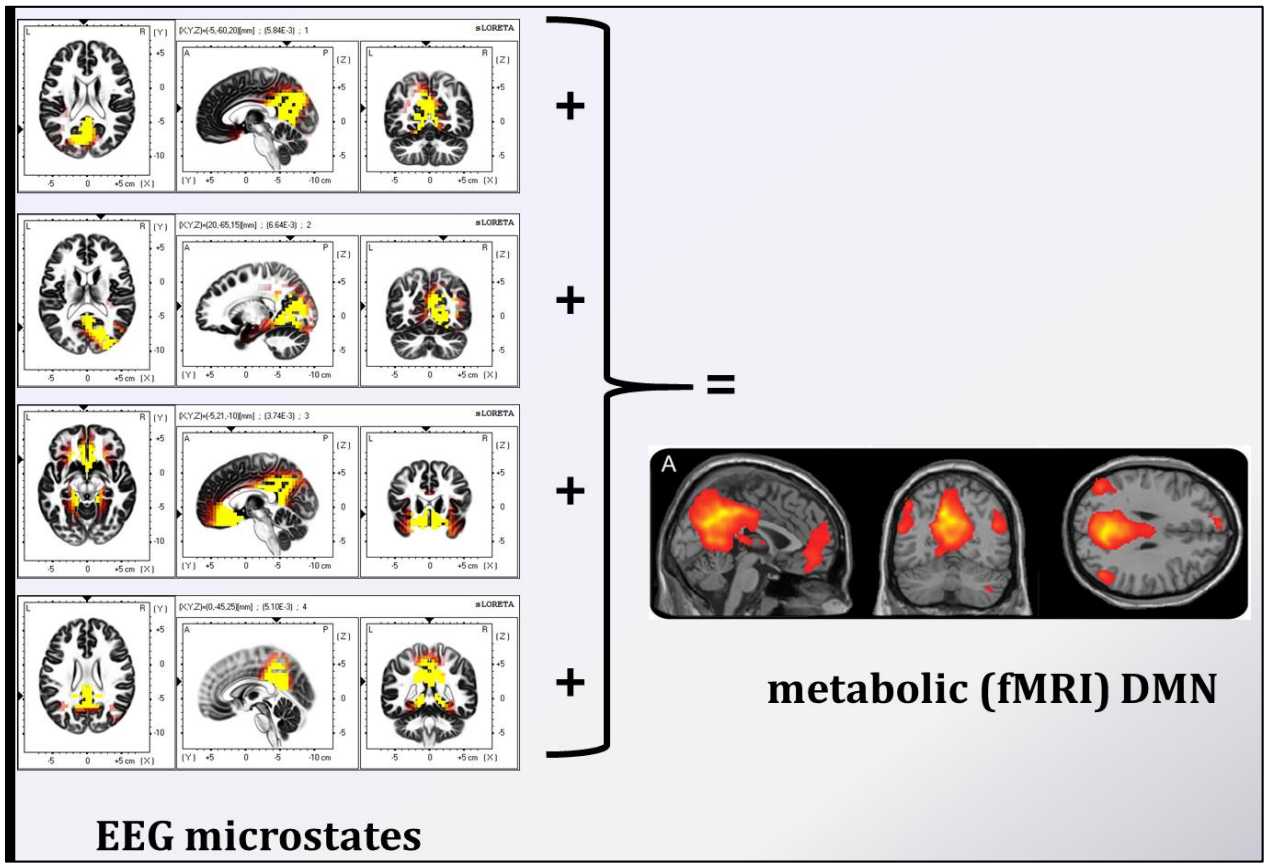

Figure 3: Schematic representation of the notion that that metabolically determined DMN actually corresponds to a very low-pass time filtered version of the faster microstate dynamics.

**6c: Confirmatory analysis for inferring the cortical distribution of electrically active neurons underlying the four microstates**

For each participant, and for each individual microstate scalp map $\Gamma_{ji} \in \mathbb{R}^{N_E \times 1}$, with $j=1...N_S$, $i=1...N_\mu$, $N_S = 109$, $N_\mu = 4$, an eLORETA 3D image was computed, consisting of current density vector (three components) at $N_V = 6239$ voxels: $\mathbf{J}_{ji} \in \mathbb{R}^{N_V \times 3}$. This material was then used to test for zero mean values at each voxel, for each current density vector component. Thus, the number of tests was $3 \times N_V \times N_\mu = 74868$. Correction for multiple testing is based on non-parametric randomization of the maximum-statistic (see e.g. Nichols and Holmes 2002).

In this way, for each microstate and at each voxel, the maximum statistic of the three components is displayed, with the corrected threshold. This is a test for absolute activation, with no need for a reference or baseline condition.

Figure 4 shows in bright yellow the localizations of the electrically active cortical regions that generate the microstates, at corrected $p<0.01$.

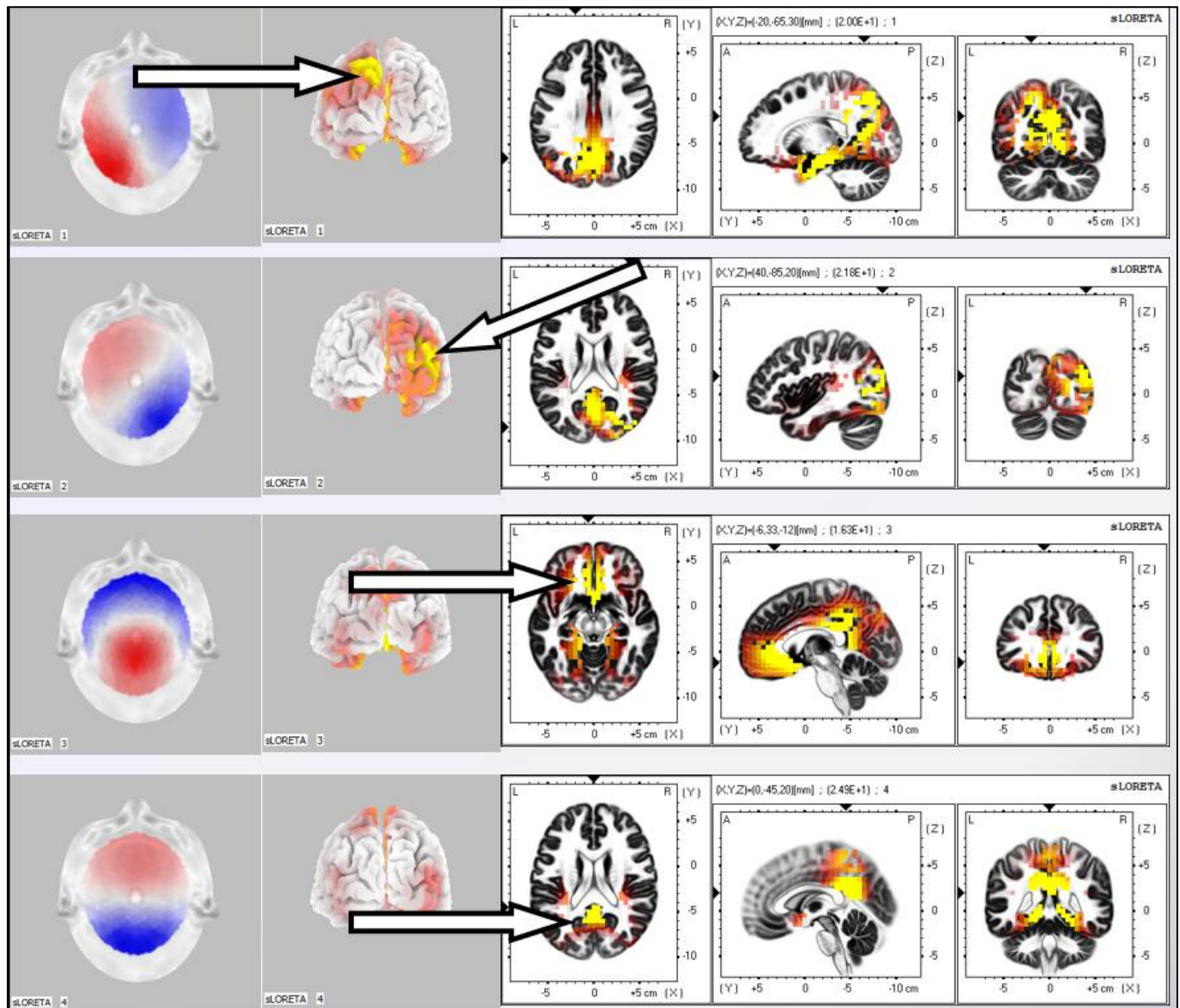

Figure 4: Left column: the four common microstate scalp maps. Middle and right columns: the corresponding cortical generators, with corrected *p*<0.01 shown in bright yellow. Arrows indicate the corresponding regions of the metabolically determined DMN.

The results in Figure 4 show that the microstates of the resting state are generated in the following cortical regions:
1: Left occipital/parietal + posterior cingulate
2: Right occipital/parietal + posterior cingulate
3: Anterior cingulate + posterior cingulate
4: Posterior cingulate

These confirmatory results once more support the following notions regarding the metabolically determined DMN:
1. The four DMN regions do not work together, as asserted in the literature.
2. They seemingly work together, due to low time resolution of metabolic imaging.

More importantly, these results, in their own right, show that:
1. The posterior cingulate is constantly activate. It is common to all four microstates.
2. During EC REST, the posterior cingulate is an extreme functional hotspot.

3. Moreover, the posterior cingulate is a central hub, linking all other areas.

### 6d. Effective connectivities in the resting brain (i.e. the resting state connectome at low spatial resolution)

Functional connectivity is of central importance in understanding brain function. For this purpose, the multiple time series of electric cortical activity of the microstates can be used for assessing the strength, directionality, and spectral characteristics (i.e., which oscillations are preferentially transmitted) of the cortical connections that are mediated by the microstate transitions.

The *microstate time series*, for each participant, denoted as $\mathbf{c}_{jt} \in \mathbb{R}^{N_\mu \times 1}$, are available from Eq. 10. In particular, these time series contain information on the dynamic behavior and interactions and transitions between the cortical generators of the microstates (shown in Figure 4).

In this study, use is made of the isolated effective coherence (iCoh) (Pascual-Marqui et al 2014) for characterizing effective cortical connectivity between the sets of cortical regions that generate the microstates. This measure has its roots in the work of Akaike(1968), Granger (1969), and Baccala and Sameshima (2001) on causality in time series.

The iCoh, denoted as $\kappa_{i \leftarrow j}(\omega)$, gives the strength of the direct influence of time series "*j*" on "*i*" at frequency ω. The term "direct influence" signifies that all possible indirect paths from "*j*" to "*i*" through other measured time series, are discounted.

The iCoh is a genuine partial coherence obtained under a multivariate autoregressive model, followed by setting all irrelevant associations to zero, other than the particular directional association of interest. The iCoh is distinct and different the partial directed coherence (PDC) of Baccala and Sameshima (2001). In Pascual-Marqui et al (2014) it is shown how the PDC can give incorrect information about the strength of a connection, and incorrect information on its spectral characteristics; and it is shown how the iCoh solves this problem.

The iCoh was calculated for each participant separately, using each ones *microstate time series* ($\mathbf{c}_{jt} \in \mathbb{R}^{N_\mu \times 1}$ from Eq. 10). For each participant, this consists of an effective directional connectivity matrix $\mathbf{K}(\omega) \in \mathbb{R}^{N_\mu \times N_\mu}$, in general non-symmetric, with zeros on the diagonal. The element in row "*i*" and column "*j*" is precisely $\kappa_{i \leftarrow j}(\omega)$. The iCoh was evaluated in the discrete frequency range from 2 to 20 Hz. The multivariate autoregression model used for estimating the iCoh was of order 7.

The z-transformed iCoh values (using 109 individual iCohs) are displayed in detail in Figure 5. The causal influence is defined from column to row (i.e. sender to receiver). The vertical axis corresponds to z-scores, and the horizontal axis to frequency, from 2 to 20 Hz. Indicated in color red are the direct causal connections with z-score higher than 15.

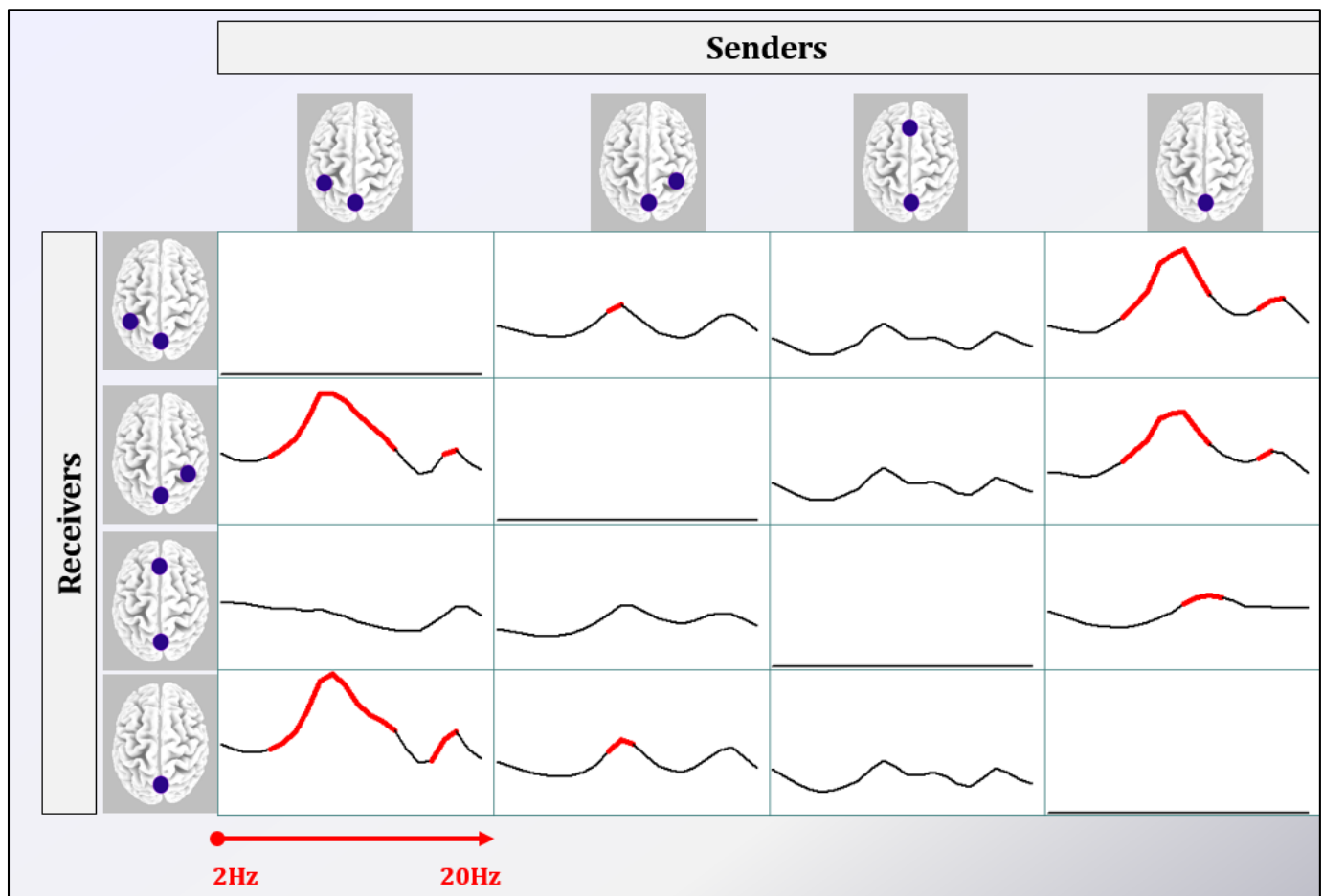

Figure 5: z-transformed iCoh values (using 109 individual iCohs) are displayed. The causal influence is defined from column to row (i.e. sender to receiver). The vertical axis corresponds to z-score, and the horizontal axis to frequency, from 2 to 20 Hz. Indicated in color red are the direct causal connections with z-score higher than 15.

A schematic representation of the results are summarized in Figure 6, which shows the strongest effective connections (z-score higher than 15) of the microstate generators. The results confirm the role of the posterior cingulate as an important in sequential feedback with the left and right parietal/occipital regions, but driving in a unidirectional manner the anterior cingulate. Furthermore, not only alpha oscillations play a role in the resting state connectome. Higher frequency oscillations (beta) are also involved in the transfer of information, despite the dominance of alpha oscillatory activity in the activity signals.

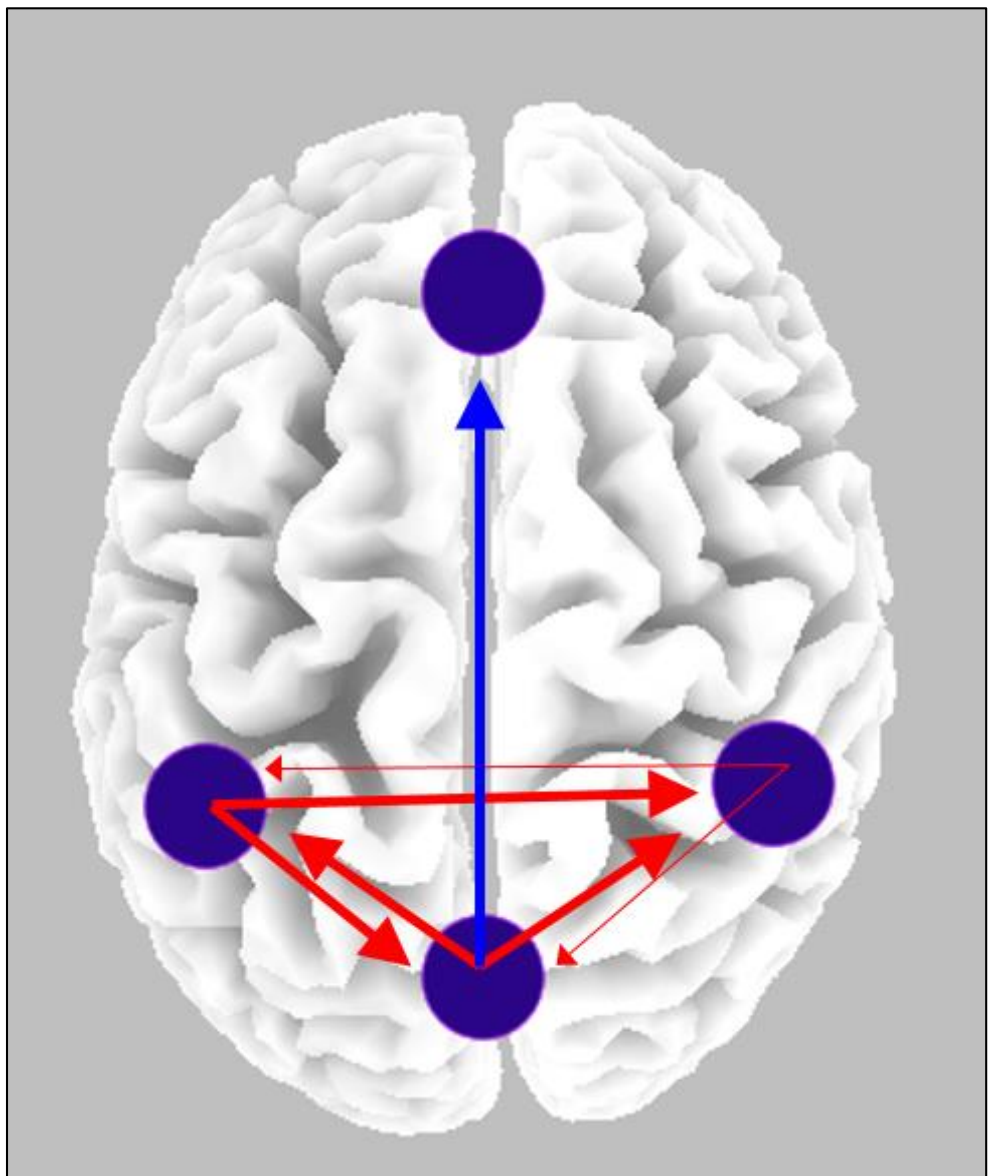

Figure 6: This is a schematic summary of the results in Figure 5, corresponding to the strongest effective connections (z-score higher than 15) of the microstate generators. This is a top view of the cortex, left is left. Large blue circles indicate the main generators of the microstates. RED arrows correspond to directional, direct information transfer of alpha (low and high) and middle beta oscillations. The blue arrow corresponds to the directional, direct information transfer of high alpha and low beta oscillations.

## 7. Summary

This study uses EEG recordings during awake, eyes closed, resting state. These are the same conditions used in "resting state network" (RSN) studies that are very common in the PET and fMRI literature. In the EEG case, the images of electric neuronal activity of the cortex have extremely high time resolution (but low spatial resolution), whereas fMRI and PET have extremely low time resolution.

It is therefore almost trivial to expect that the main metabolic network of the resting brain, namely the default mode network (DMN), is a time averaged version of the electrical networks. This is confirmed in this study, under the microstate model.

However, the results of this study are not trivial, since they clarify and shed light on the actual detailed mechanism that gives place to the metabolically observed DMN. The precise interactions and oscillatory characteristics between the major cortical areas are now evident.

It should be noted that these results do not provide any proof that there is an actual relation between the EEG microstates and the metabolic DMN. In order to study and test if this is true, it would be necessary to manipulate the brain state, and to compare if the changes in microstate connectivities and metabolic networks are correlated.